# Establishment of earth tides effect on water level fluctuations in an unconfined hard rock aquifer using spectral analysis

**JC. Maréchal[1], M.P. Sarma[2], S. Ahmed[3] and P. Lachassagne[4]**

[1] BRGM (Bureau de Recherches Géologiques et Minières – French Geological Survey), Indo-French Centre for Groundwater Research, NGRI, Uppal Road, Hyderabad – 500 007, India (*Corresponding author*)

[2] NGRI (National Geophysical Research Institute), Uppal Road, Hyderabad – 500 007, India

[3] NGRI (National Geophysical Research Institute), Indo-French Centre for Groundwater Research, NGRI, Uppal Road, Hyderabad – 500 007, India

[4] BRGM (Bureau de Recherches Géologiques et Minières – French Geological Survey), Service EAU, Unite RMD, 1039 avenue de Pinville, F-34000 Montpellier, France

**Abstract**: Short-interval water level measurements using automatic water level recorder in a deep well in an unconfined crystalline rock aquifer at the campus of NGRI, near Hyderabad shows a cyclic fluctuation in the water levels. The observed values clearly show the principal trend due to rainfall recharge. Spectral analysis was carried out to evaluate correlation of the cyclic fluctuation to the synthetic earth tides as well as groundwater withdrawal time series in the surrounding. It was found that these fluctuations have considerably high correlation with earth tides whereas groundwater pumping does not show any significant correlation with water table fluctuations. It is concluded that earth tides cause the fluctuation in the water table. These fluctuations were hitherto unobserved during manual observations made over larger time intervals. It indicates that the unconfined aquifer is characterised by a low porosity.

**Keywords**: earth tide, water level, hard rock aquifer, groundwater, spectral analysis

## INTRODUCTION

Water levels in aquifer are an important parameter in groundwater hydrology and a careful and detailed analysis of its spatio-temporal variation reveals useful information on the aquifer system. Among various causes affecting the groundwater levels are groundwater withdrawals, rainfall recharge, evapo-transpiration, interaction with surface water bodies etc. Ocean tides are also known to affect the groundwater fluctuation in the coastal aquifers.

The water levels measured in a well located in an unconfined hard rock aquifer, located far away from the sea are characterised by cyclic fluctuations. Knowing that fluctuations due to evapo-transpiration cycles appear only in very shallow aquifers and are consequently not relevant in the studied case, these fluctuations can be due to two different factors. The first one is anthropogenic. It is well known that the groundwater withdrawal from an aquifer or from a field of wells induces water level decline creating a cone of depression depending, among other parameters, on the aquifer hydrodynamic parameters and geometry. But, after the pumping is stopped, the level starts coming up due to recuperation to attain equilibrium. This gives a sort of cyclic fluctuation if the pumping follows a regular interval and the levels are recorded continuously.

The second cause able to induce daily water level fluctuations is the earth tides. Actually, the effect of earth tides has been be observed in the groundwater level fluctuation of an aquifer, if monitored continuously or at a shorter interval [1, 2, 3]. The dilatation of the Earth due mainly to the position of Moon and Sun induces measurable water level fluctuations in the well. The effects of earth tides can be observed in "most wells completed in a well-confined aquifer", according to [1], and in aquifers that have "low porosity" and are "relatively stiff", according to [4].

A time series of water levels, gravity field fluctuations due to earth tides and water pumping cycles have been studied using spectral analysis to identify the principal cause affecting the water levels in the studied well.

## WATER LEVEL OBSERVATIONS

The observations of water level fluctuations have been taken from a well located in the NGRI-campus in Hyderabad (figure 1), in a hard-rock region, using a data logger from 1$^{st}$ June 2000 to 12$^{th}$ June 2000 with a time interval of 42'04''.

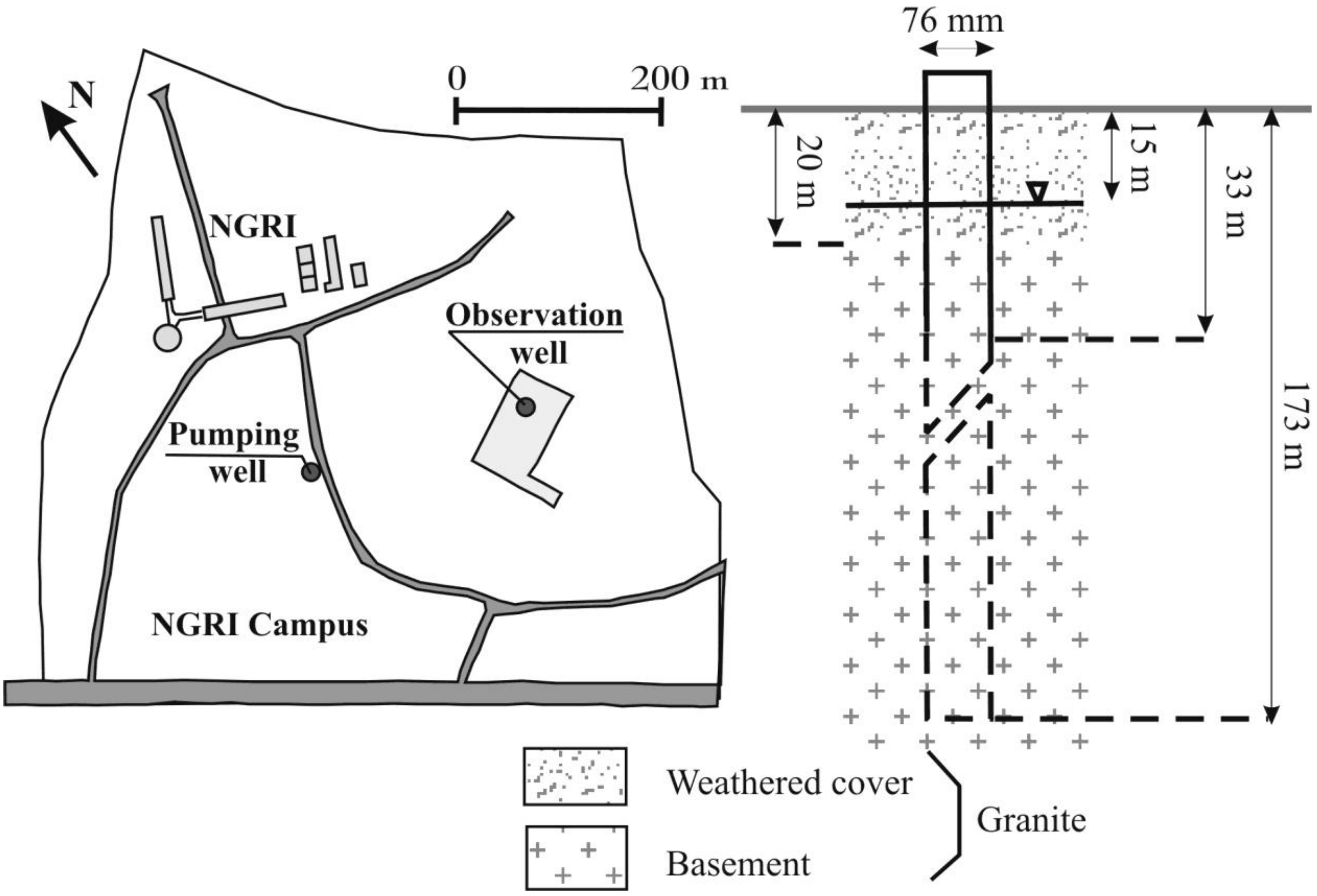

*Figure 1: location map and schematic geological cross-section of the observation well*

The well (location: 78° 33’ 03.2’’; 17° 25’ 02.1’’) was drilled in granite up to 173m cutting across a number of minor/major fractures after a weathered cover of about 20m. The groundwater depth is about 15 metres before the Monsoon and 9 metres after.

The water level (water level above the pressure probe) has an increasing trend corresponding to the effect of recharge from the rainfall (figure 2). The moving average (on 50 values, averaging window of 35 hours duration) is slowly increasing between the first and the $5^{th}$ June and the same has greater slope after the $5^{th}$ June. After a relatively dry month of May with only two days of rain, high rainfall of about 72mm occurred on $4^{th}$ June at the meteorological station of NGRI, a few hundreds of meters from the well. After that, a couple of more rainfall events appeared during the measurement period.

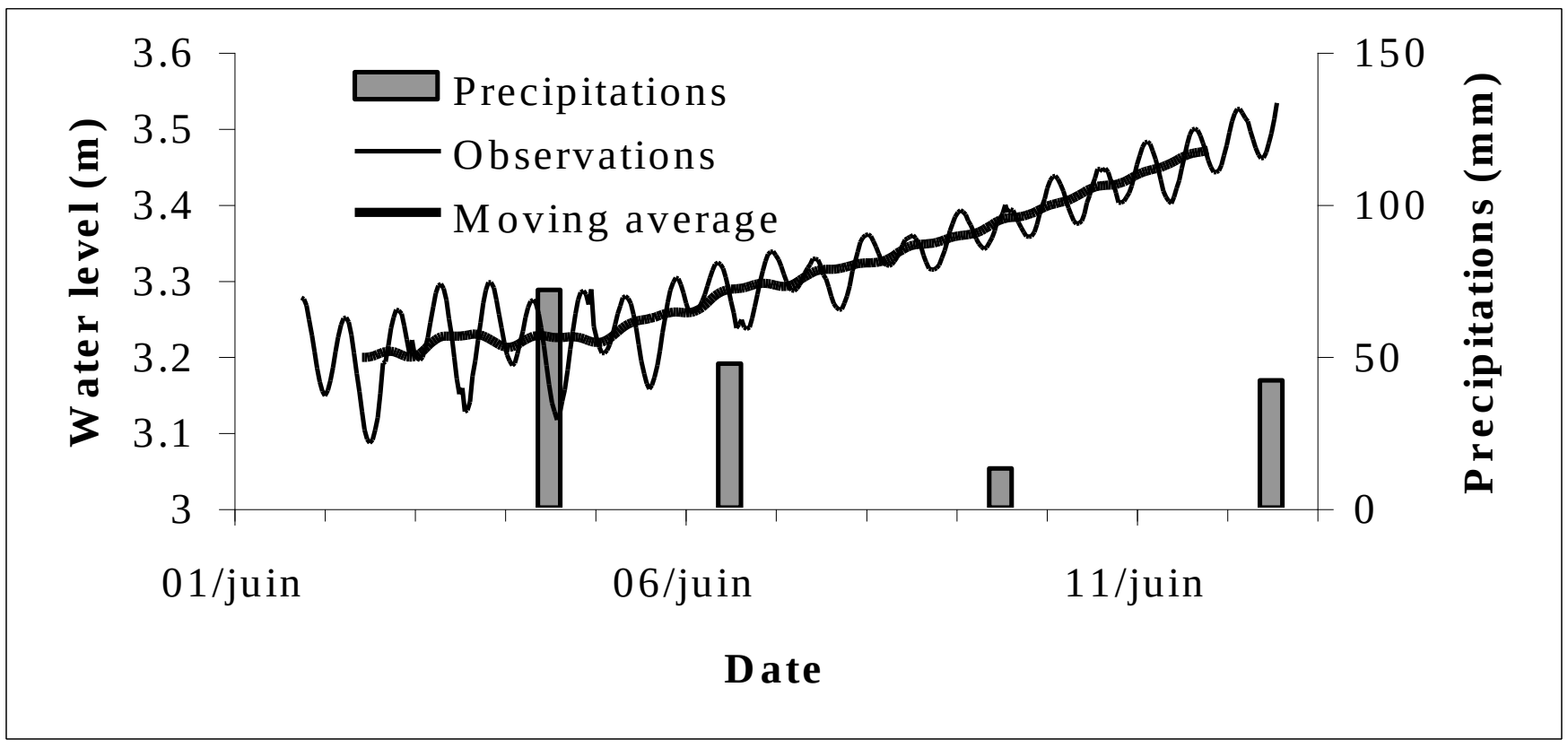

*Figure 2: Water levels and precipitation during the observation period*

A typical annual variation of water level with rise during the Monsoon and decline in the remaining period with an inlet of the study period is shown in Figure 3. It appears just at the beginning of the water level increase due to the monsoon recharge.

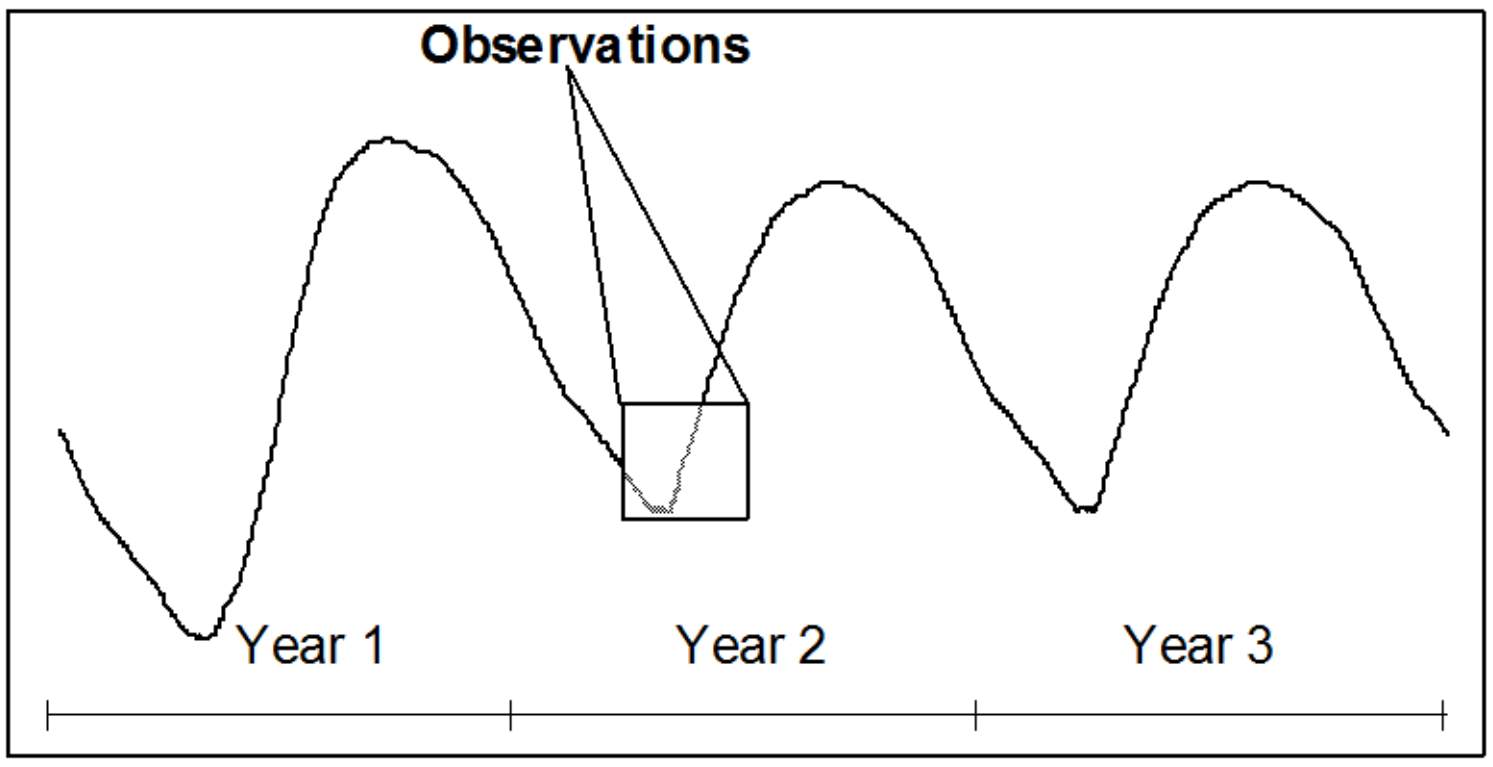

*Figure 3: Schematic diagram showing time-evolution of groundwater level in a monsoon region*

## EARTH TIDES AND WATER PUMPING

Earth tides are the result of a visco-elastic deformation of the Earth under the action of gravitational pull of Moon and Sun. Among the 386 existing tide waves, only high amplitude waves have an effect on the aquifers. Melchior [3] indicates that five large main waves are responsible for almost 95% of water level fluctuations observed in wells. These waves can be divided into two groups: tesseral waves of daily period and sectorial waves of semi-daily period (table 1).

*Table 1: Origins, periods and frequency of the principal tidal fluctuations*

| Code | Frequency (degrees/hour) | Period (d) | Origin |
|---|---|---|---|
| $M_f$ | 1°098’033 | 13.66 | Lunar, variation of declination |
| $O_1$ | 13°934’036 | 1.07 | Lunar, main term – tesseral |
| $K_1$ | 15°041’069 | 0.99 | Lunisolar, main term - tesseral |
| $N_2$ | 28°439’730 | 0.53 | Lunar, main term (orbital ellipsoid) – sectorial |
| $M_2$ | 28°984’104 | 0.52 | Lunar, main term - sectorial |
| $S_2$ | 30°000’000 | 0.42 | Solar, main term |

These tides induce a cubic dilatation of Earth which is responsible for water level fluctuations in wells. Gravity field fluctuations due to synthetic earth tides have been computed at the well site in Local Standard Time (Time Zone : 5 East) with a special code taking into account the main tidal components.

The only water pumping present in the neighbourhood of the observation well is located at 215 m to the Southwest (figure 1). Groundwater pumping is of variable duration everyday in the well.

Figure 4 illustrates the comparison between observed water level fluctuations, computed earth tides effect on gravity field (μGals) and water pumping periods in local time (GMT + 5:30). The cycles of earth tides seem to have the same frequency than water level fluctuations. A local maximum of gravity corresponds to a local maximum of water level. On the other hand, pumping does not seem to be correlated with water level: some pumping periods are characterised by an increasing water level while some off periods are characterised by a declining level.

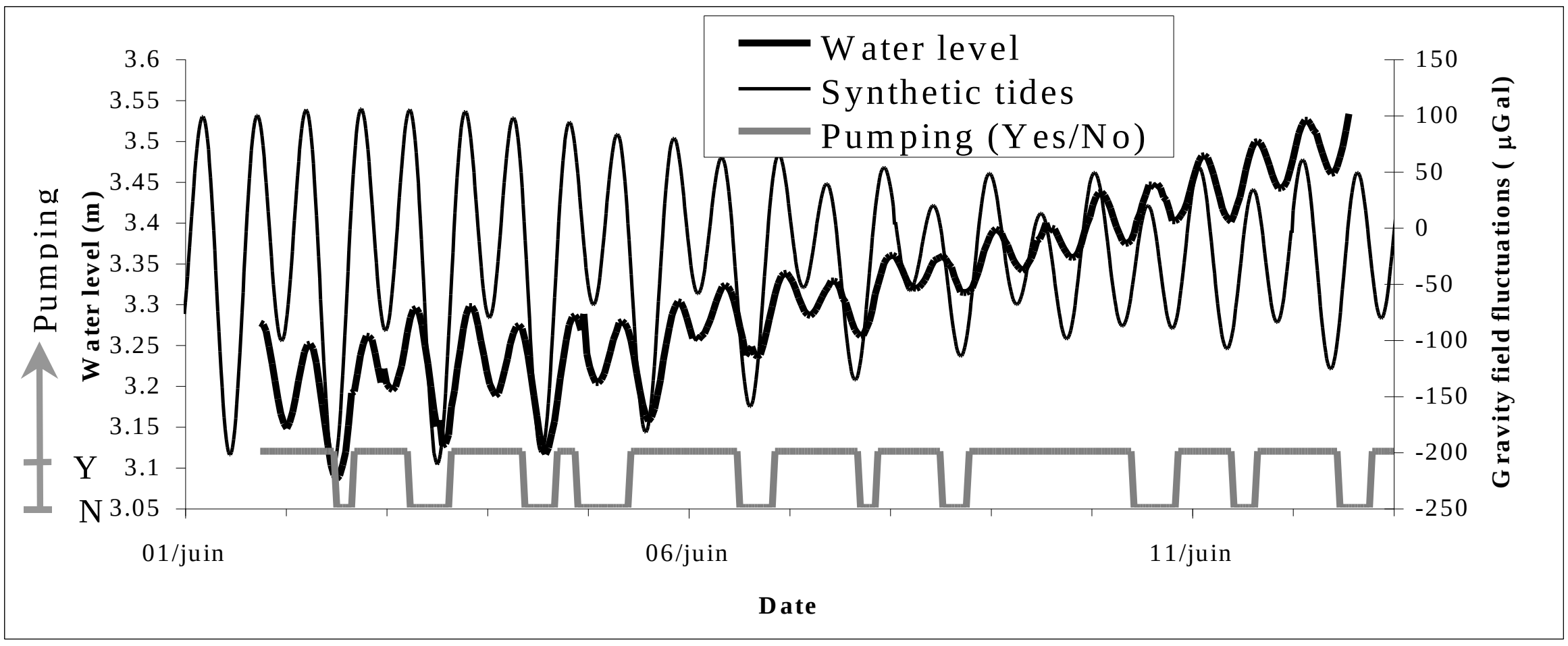

*Figure 4: Gravity field fluctuations due to synthetic tides, water level fluctuations and water pumping (Yes or no).*

During the observation period, the detailed analysis shows that maximum earth tides fluctuations between 2nd and 3rd of June correspond to maximum water level fluctuations with an amplitude of 10 centimetres and the minimum of earth tide fluctuations during the 8th of June corresponds to a minimum of water level fluctuation with an amplitude of less than 3 centimetres. The amplitude of water level fluctuations was used by [5] applying the theory of [1] to determine the storage coefficient of the aquifer.

## SPECTRAL ANALYSIS

A preliminary observation of the studied signals tends to show that the water level fluctuations are better correlated with earth tides than with water pumping. This statement is verified by spectral analysis.

A filter must be applied on the water level data to enhance interesting components and remove the ones that hide the cycles of interest. In the studied case, the trend corresponding to the monsoon recharge needs to be removed as seen at figures 2 and 3. As proposed by [5] according to [6], the trend can be removed using a first order differencing calculating the punctual increase of the time data series. Given X(t) a time data series, the differentiated Y(t) is computed as following:

$$Y(t) = X(t-1) - X(t)$$

This method conserves the structure and the amplitude of phenomena while removing the trend [5]. The data used below have been filtered using this method.

The computation of the power spectrum allows identification of the frequencies of a periodic signal. For a given signal, the power spectral density is high for the frequencies that characterise this signal and is low for other frequencies. Thus, as an example, the power spectral density of a periodic signal with three cycles at frequencies, say $f_1$, $f_2$ and $f_3$ will be characterised by three peaks at the frequencies $f_1$, $f_2$ and $f_3$. The relative power spectral density was computed for the synthetic earth tides, filtered water level observations and water pumping data (Figure  5). While the power spectral density of water pumping is different, that of water levels and earth tides are similar with two peaks appearing at frequencies $f_1$ and $f_2$. They are described in Table 2. The periods of these cycles correspond exactly to the daily tesseral waves $O_1$ and $K_1$ and to the semi-daily sectorial waves $N_2$ and $M_2$.

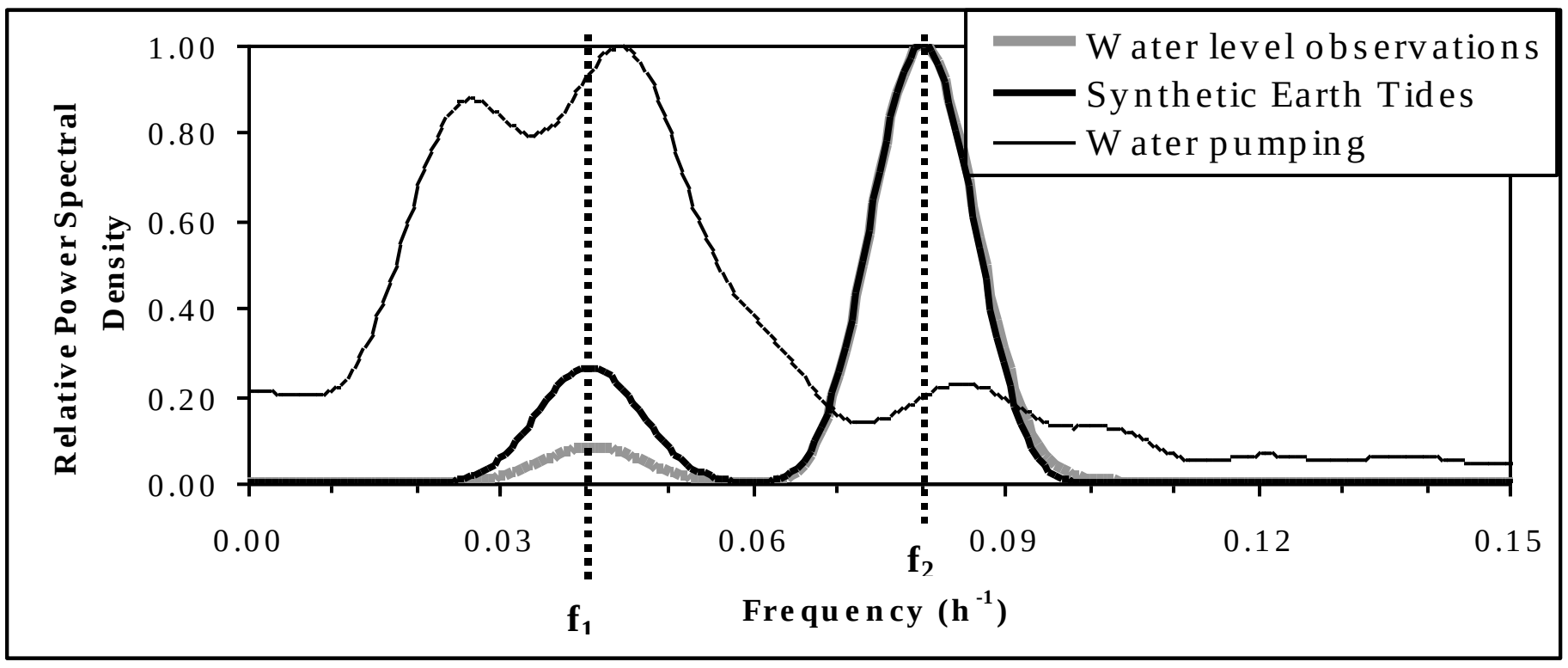

*Figure 5: Relative power spectral density of synthetic earth tides, pumping cycles and filtered water levels*

| *Frequency* | *Period* | *Earth tide* |
|---|---|---|
| $f_1 = 0.04109\ h^{-1}$ | $P_1 = 24.34$ hours (1.01 day) | $O_1$, $K_1$ |
| $f_2 = 0.08009\ h^{-1}$ | $P_2 = 12.48$ hours (0.52 day) | $N_2$, $M_2$ |

*Table 2: Characteristics of observed frequencies on the signals*

The amplitude function expresses, for each frequency, the magnitude of the input-output relation. Usually, only relations between input and output for frequencies where amplitude of covariance is high can be interpreted. For other frequencies, a low covariance indicates that there is no input-output relation. In this case (figure 6), the high value of amplitude at frequencies $f_1$ and $f_2$ confirms that there is a relation between earth tide (as input) and water level (as output) fluctuations. This is not the case for pumping and level fluctuations.

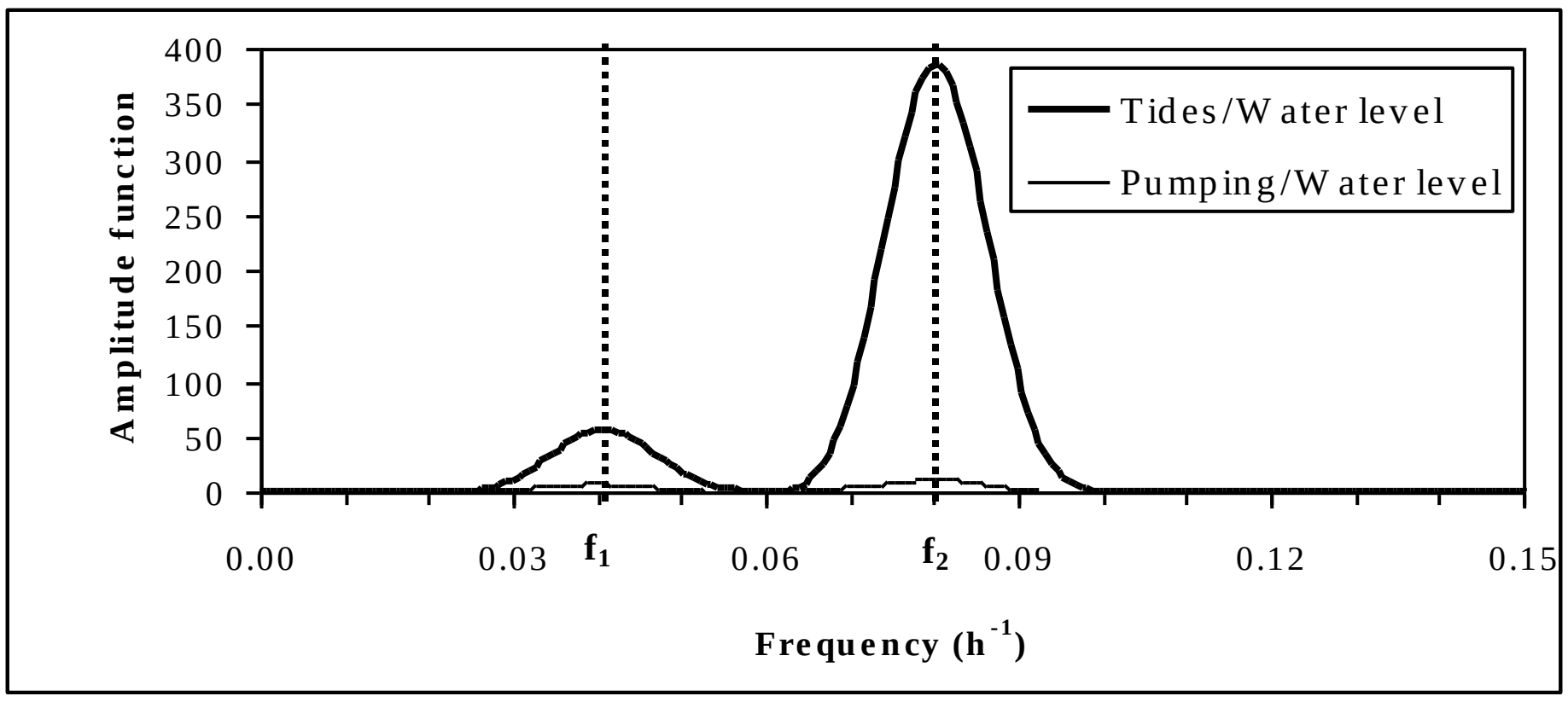

*Figure 6: Amplitude function between tidal fluctuations and filtered water levels, and between pumping cycles and filtered water levels.*

Coherency describes the degree of relation between two signals of same period. The square of coherency lies between 0 and 1. When $|Csr|^2 = 1$, for a given frequency $f = f_0$, then, there is a linear relation between both signals (at the frequency $f = f_0$). Figure 7 shows that for

earth tides and water levels, $|Csr|^2 = 1$ at frequencies $f_1$ and $f_2$. This indicates that earth tides and water levels signals not only have the same frequencies, but are linearly related at these frequencies. The coherency between water pumping and levels is low at all the frequencies. There is no direct linear relation between these signals.

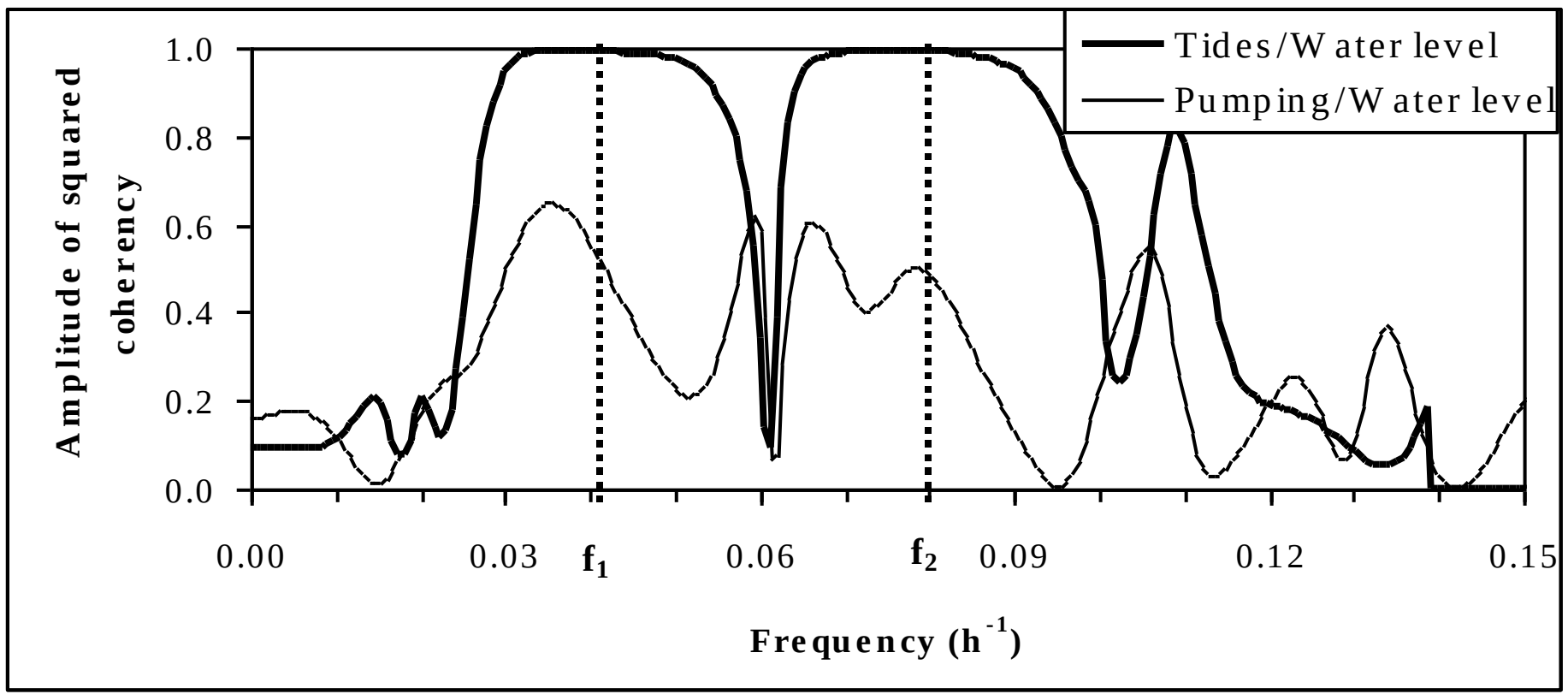

*Figure 7: Coherency between tidal fluctuations and filtered water levels and between water pumping cycles and levels.*

The cross-correlation function represents the inter-relationship between the input and output series. The time lag at which the maximum cross-correlation occurs determines the stress transfer velocity of the system. The cross-correlation diagram between tidal fluctuations and water levels shows that the correlation at the origin is positive (figure 8): a local maximum of gravity corresponds to a local maximum of water level: high Earth gravity (due to lower Moon or Sun attraction) induces a contraction of the Earth and a water level increase due to decrease of porosity. A maximum positive correlation is found after about one hour. This is the delay in the reaction of the aquifer to the earth tide depending on hydraulic conductivity and storage coefficient of the aquifer [7]. Various scientific approaches exploit this dependence to determine the hydraulic parameters of an aquifer using the response of water level to earth tides [5, 7, 8, 9].

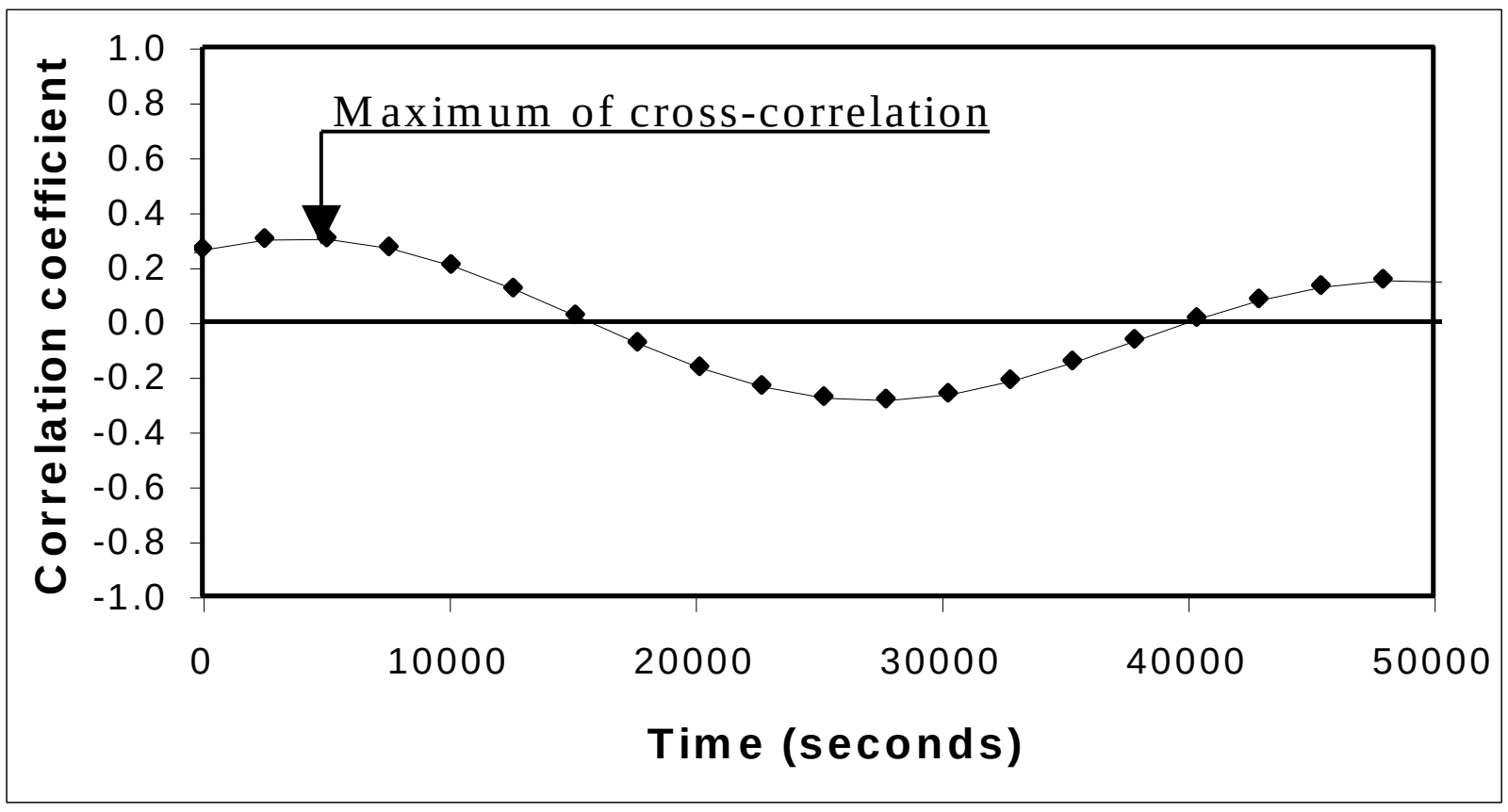

*Figure 8: Cross-correlation function between gravity fluctuations due to synthetic earth tides and water levels*

## CONCLUSIONS

The detailed study of water level fluctuations measured in a well located in an hard-rock aquifer shows that earth tides and not water pumping are responsible for observed daily and semi-daily period fluctuations. Frequencies of synthetic earth tides and water level observations signals are similar. Tesseral waves of daily period ($O_1$, $K_1$) and sectorial waves of semi-daily period ($N_2$, $M_2$) are identified as the origin of the water level fluctuations. Spectral analysis has shown that there is a direct linear relation between earth tides and water level fluctuations in the studied well. The observation of such fluctuations in this hard-rock aquifer apparently unconfined implies that the aquifer is characterised by a low porosity.

## ACKNOWLEDGMENTS

The authors thank Dr JL Pinault from BRGM (French Geological Survey) who developed the software TEMPO used for the processing of signals. They also thank Mr T.R.M. Prasad for his advise on computation of synthetic tides.

**REFERENCES**

1. Bredehoeft, J.D., *J. Geophys. Res.*, 1967, **72**(12), 3075-3087.

2. Bovardson, G., *J. Geophys. Res.*, 1970, **75**(14), 2711-2718.

3. Melchior, P., *The Tides of the Planet Earth*. Pergamon, Paris, 1978, 609 p.

4. Rojstaczer, S. and Agnew, D.C., *J. Geophys. Res*., 1989, **94**(B9), 12403-12411.

5. Marsaud, B., Mangin, A. and Bel, F., *Journal of Hydrology*, 1993, **144**, 85-100.

6. Box, G.E.P. and Jenkins, G., *Time Series Analysis: Forecasting and control*, Holden Day, San Francisco, 1976, 575 p.

7. Hsieh, P.A., Bredehoeft, J.D. and Farr, J.M., Eos, Transactions, American Geophysical Union, 1985, **66**(46), pp. 891.

8. Mehnert, E., Valocchi, A.J., Heidari, M., Kapoor, S.G. and Kumar, P., *Groundwater*, 1999, **37**(6), 855-860.

9. Ritzi, R.W., Sorooshian, S., Hsieh, P.A., *Water Resources Research*, 1991, **27**(5), 883-893.